\def\ueber#1#2{{\setbox0=\hbox{$#1$}%
  \setbox1=\hbox to\wd0{\hss$\scriptscriptstyle #2$\hss}%
  \offinterlineskip
  \vbox{\box1\kern0.4mm\box0}}{}}
\def \b {\vec}
\def \mathrm{\rm}

\def\parts{\partial_{\sigma}}

\def\partp{\partial_{\Phi}}

\def\partt{\partial_{t}}
\def\lsim{\stackrel{<}{{}_\sim}}
\def\gsim{\stackrel{>}{{}_\sim}}
\def\calP{{\cal P}}
\def\e{{\mathrm{e}}}

\def\d{{\mathrm{d}}}

\catcode`@=11
\def\gsim{\mathrel{\mathpalette\@versim>}}
\def\@versim#1#2{\lower0.2ex\vbox{\baselineskip\z@skip\lineskip\z@skip
       \lineskiplimit\z@\ialign{$\m@th#1\hfil##$\crcr#2\crcr\sim\crcr}}}
\catcode`@=12
\catcode`@=11
\def\lsim{\mathrel{\mathpalette\@versim<}}
\def\@versim#1#2{\lower0.2ex\vbox{\baselineskip\z@skip\lineskip\z@skip
       \lineskiplimit\z@\ialign{$\m@th#1\hfil##$\crcr#2\crcr\sim\crcr}}}
\catcode`@=12
\def\etal{{\sl et al.~}}
\def\v{{\cal V}}
\documentstyle[12pt]{article}
\advance\textheight by \topskip
\begin{document}
\thispagestyle{empty}

\newcommand\be{\begin{equation}}
\newcommand\ee{\end{equation}}

\today \hspace{2.8in} {\raggedleft Submitted to Phys. Rev. D \par}
\vskip 2cm
\begin{center}
{\huge How typical is General Relativity}
\vskip 0.3cm
{\huge in Brans-Dicke chaotic inflation?}
\vskip 1.5cm
{\large \sc Mikel Susperregi}\footnote{\it m.susperregi@qmw.ac.uk}
\\
\vskip 1cm

{\it Astronomy Unit, School of Mathematical Sciences,\\
Queen Mary \& Westfield College, London E1 4NS, United Kingdom\\}

\end{center}

\vskip 1.5 cm

\begin{center}
{\large\bf Abstract}
\end{center}
\begin{quotation}
\vskip -0.4cm
\noindent General Relativity is recovered from Brans-Dicke 
gravity in the limit of large $\omega$. In this article we 
investigate theories of Brans-Dicke gravity with chaotic 
inflation, allowing for either a constant or variable value 
of $\omega$, known as {\it extended} and {\it hyperextended} 
inflation respectively. The main focus of the paper is placed 
on the latter. The variation $\omega$ with respect to 
the Brans-Dicke field is based on higher-order corrections 
analogous to those of the dilaton field in string theory, 
following the simple principle that the Brans-Dicke and metric 
fields decouple asymptotically. The question addressed is 
whether a large value of $\omega$ is predominant in most 
regions of the universe, which would lead to the conclusion 
that a typical region is then governed by General Relativity. 
In these theories we find that it is possible to construct 
inflaton potentials that drive the evolution of the Brans-Dicke 
field to an appropriate range of values at the end of inflation, 
such that a large $\omega$ is indeed typical in an average region 
of the universe. However, in general this conclusion does not 
hold and it is shown that for a wide class of inflaton potentials 
General Relativity is not a priori a typical theory.

\end{quotation}

\newpage

\section{Introduction}

The inflationary paradigm in cosmology is well understood and 
it provides a suitable framework to describe the early evolution of 
the universe \cite{inflation-general}. Inflation has taken many 
forms and undoubtedly the key to its success 
is its ability to adopt changes. Although initially 
a inequivocal prediction of inflationary models was a density 
parameter $\Omega=1$, presently a suitable choice of the 
scalar field potential has been shown to 
yield open ($\Omega<1$) \cite{open-inflation} or even closed 
($\Omega>1$) \cite{closed-inflation} inflation. 
The uncertainty in the determination 
of $\Omega$ from various dynamical data, such as velocity flows and 
other large-scale data  
\cite{omega}, enables us to produce a large class 
of inflationary models that are consistent with  
constraints from cosmic microwave background (CMB) fluctuation maps. 
Hopefully, in the next few years, the values of cosmological 
parameters such as $\Omega$, the cosmological constant 
$\Lambda$, the Hubble parameter $h$ and 
the initial spectral index $n$ will be pinned down to considerable 
accuracy with CMB data from the forthcoming 
{\it Planck surveyor} and ground-based CMB experiments 
\cite{CMB}. 
A better determination of $\Omega$ will finally narrow down 
the family of inflationary models, which technically translates 
into a better knowledge of the inflaton potential, and ultimately 
on the particle physics involved. \\
\\
Chaotic inflation theories in particular, are successful in 
explaining the generation of primordial density perturbations 
\cite{inflation-general,chaotic-inflation,structure-extended} 
and they additionally generate a quantum cosmological scenario,  
as we will discuss below, that one can compare with other approaches 
in quantum cosmology \cite{wavefunction}. 
The scalar field evolves from initial 
large values and after a brief period of inflation, a homogeneous 
region becomes subdivided in further regions where the inflaton 
field takes a wide range of values. The scalar field reaches the end 
of inflation in some of these regions, whereas others continue 
inflating and are in turn subdivided in further regions where, 
as before, the field takes a wide range of values. 
The process of inflation is {\it eternal} and 
{\it self-reproducing}, in the sense that a homogeneous 
subregion is subdivided in inflating and non-inflating regions 
in much the same way as regions at earlier and later stages do 
\cite{self-reproduction}. At any given value of the radius 
of the universe, inflation still takes place and the values 
of the fields are thus given in terms of a probability distribution 
$P(\sigma)$ over an ensemble of regions of the universe that 
is governed by the stochastic equation motion of the scalar 
field \cite{stochastic}. In this  
quantum cosmological scenario  
$P(\sigma)$ tells us the likelihood of a certain region 
being at a certain stage of inflation. These theories 
are unable to make 
definite predictions for any given region due to the stochastic 
nature of inflation. The self-reproducing scenario also tells 
us that although constraints on the age of the universe obtained 
via stellar evolution, element abundances, etc, will ultimately 
yield an estimate of the time elapsed since our region of the 
universe stopped inflating, those constraints will not 
give us any information of the age of the universe as a whole. 
The property of self-reproduction permits 
the possibility of a universe that extends to the infinitely 
remote past. On the other hand, it has been shown that, 
given a homogenous region where inflation sets on at 
an initial time, then the distribution 
$P(\sigma)$ quickly approaches a stationary regime 
\cite{stationary}, and is therefore solely dependent 
on the initial conditions. \\
\\
In the case of several scalar fields, the interplay between 
the evolution of a scalar field and the inflaton field influences 
the course of inflation, and the {\it self-reproducing universe} 
is then described in terms of the joint probability distributions 
$P(\sigma;\Theta)$, where $\Theta$ denotes scalar fields that are 
coupled to inflation.\footnote{A more coarse description would 
be achieved by computing the partial distribution for  
$\sigma$, if one is not interested in the values of the other 
scalar fields, by computing the integral $\int \d\Theta\, 
P(\sigma;\Theta)$.} The joint evolution of the fields 
produces a quantum cosmological scenario where the fields evolve 
over a self-similar ensemble of regions. One such scalar 
field is the Brans-Dicke (BD) field $\Phi$, first introduced 
in the context of more generalized theories of gravity 
\cite{brans-dicke}. \\
\\
BD gravity was initially motivated by Mach's principle 
and a dimensional argument due to Dennis Sciama \cite{sciama} 
that relates the magnitude of $G$, the horizon radius $H^{-1}$  
and the total mass $M$ within the horizon via $GMH\sim 1$.  
The BD field determines the magnitude of $G$ 
(and therefore $M_P$) and is slowly-varying over horizon 
scales and its coupling coefficient to the curvature is 
denoted by $\omega$. In string theory, the BD field arises 
for the one-loop string effective action in the form of 
the {\it dilaton} field \cite{string}, 
the equivalent of a BD field with variable $\omega$. 
Inflationary cosmology can naturally adopt BD gravity 
and the result is the so-called {\it extended inflation} 
(constant $\omega$) that was first suggested by 
La and Steinhardt \cite{extended}. Extended inflation 
revives the old inflation scenario \cite{old} in that a first-order 
transition is responsible for inflation, but the addition of 
the BD term in the action terminates the phase transition 
with a bubble spectrum that is consistent with structure formation.  
However, limits on CMB anisotropies 
suggest that extended inflation can only work for 
$\omega\leq 25$ \cite{constraints}, in contrast with 
time-delay experiments \cite{time-delay} that set a lower 
bound $\omega\gsim 500$. This incompatibility was 
reconciled by introducing {\it hyperextended inflation} 
\cite{hyperI,hyperII}, which allows for a variable $\omega$. 
In spite of the apparent incompatibility of the CMB data with 
extended inflation, these theories have been studied in some detail, 
following the thesis that the self-reproducing 
scenario can lead to a wide range of spectra of bubble sizes, 
depending on the initial value of the $\Phi$ field (see last reference 
in \cite{constraints} for a discussion of this argument to reconcile 
CMB data with extended inflation).  
The distributions $P(\sigma,\Phi)$ for several extended 
inflation models have been derived, the spectrum 
of density fluctuations at the end of inflation  
\cite{BD-inv-perturbations,BD-perturbations} and models 
of formation of cosmic structure \cite{structure-extended}. \\
\\
The purpose of this paper is to investigate in more detail 
some implications of hyperextended inflation and ask ourselves how 
typical General Relativity (GR) is in these theories. GR is 
an accurate theory of gravity in our neighbourhood, 
and a large value $\omega\geq 500$  
reduces BD gravity to GR. We compute the probability distribution 
of $\omega(\Phi)$ in hyperextended inflation and address the question 
of whether the most typical values of $\omega$ at the end of inflation 
in such a universe are compatible with the large values required 
to recover GR. The functional dependence of $\omega$ 
on the BD field $\Phi$ is given by higher-order corrections 
in the effective string action, following the 
{\it principle of least coupling} \cite{string}. 
This principle states that 
the field $\Phi$ will evolve in a way that in the asymptotic 
regime its coupling to matter will vanish. The principle of 
least coupling enables us therefore to have a convergent 
and well-behaved approximate form of $\omega$ so that 
we can calculate the likelihood of an arbitrary value 
of $\omega$ in any region of the universe. The conclusion 
of our analysis is that although inflaton potentials can be 
chosen so that the likeliest values of $\omega$ are driven 
towards large values compatible with GR, in general the 
reverse is not true, and it is not the case that for an 
arbitrary potential GR is a typical theory. \\
\\
This article is structured as follows. Section 2 gives 
a brief summary of the essentials of extended inflation. The 
reader familiar with this can skip this section and proceed 
directly to Section 3, where hyperextended inflation is 
studied and applied to the case of powerlaw potentials. 
In Section 3 we compute the probability distributions 
$P(\sigma,\Phi)$, volume ratios and in general we discuss 
the likelihood of physical quantities, in particular the 
probability $P(\omega)$ for some simple Ansatze of $\omega$.  
In Section 4 we discuss these results and investigate 
arguments of naturalness and typicality to look into 
the question of how typical GR is in terms of $P(\omega)$. 
In Section 5 we sum up with some conclusions.

\section{Extended inflation}

Extended inflation is governed by the action 
\cite{extended,BD-inv-perturbations}
\be \label{action}
{\cal S} = \int\! \d^4 x\,\sqrt{-g}\,
\Big[\Phi\, R-{\omega\over\Phi}(\partial\Phi)^2
-{1\over 2}(\partial\sigma)^2-V(\sigma)\Big],
\ee
where $R$ is the curvature scalar and $V(\sigma)$ the 
inflaton potential. The potential energy of the BD field 
is assumed to be zero or negligible in comparison to that 
of the inflaton field. The Planck mass is related to 
the BD field, 
\be 
M_P^2(\Phi)= 16\pi\,\Phi.
\ee
The beginning-of-inflation boundary (BoI) in (\ref{action}) 
is given by
\be \label{BoI} 
V(\sigma)=M_P^4(\Phi),
\ee
and similarly the end-of-inflation boundary (EoI) is 
marked by the condition
\be \label{EoI}
{1\over 2}\dot\sigma^2+\omega\,{\dot\Phi\over\Phi}^2
\approx V(\sigma).
\ee
The resulting equations of motion in a FRW expanding 
background read, in the slow-roll approximation, i.e. 
$\ddot\Phi\ll H\,\dot\phi\ll H^2\Phi$, $\dot\sigma^2+2\omega\,
\dot\Phi^2/\Phi\ll 2\,V(\sigma)$, 
\be \label{SR1}
{\dot\Phi\over\Phi}=2\,{H\over\omega},
\ee
\be \label{SR2}
\dot\sigma = -{1\over 3H}\, V'(\sigma),
\ee
\be  \label{SR3}
H^2(\sigma,\Phi)={1\over 6\Phi}\,V(\sigma).
\ee
As is shown in \cite{BD-inv-perturbations}, 
the following conservation law follows from (\ref{SR1})-(\ref{SR3}): 
\be  \label{conservation}
{\d\over\d t}\biggl[ \omega\Phi+\int\!\d\sigma\,{V(\sigma)
\over V'(\sigma)}\biggr]=0.
\ee
The classical trajectories of the fields are then given 
by the integrals of (\ref{conservation}). In the ($\sigma$,$\Phi$) 
plane the integral of (\ref{conservation}) is a  
parabola in the case of a powerlaw potential, a straight line 
for an exponential potential and quasi-logarithmic 
trajectories for double-well potentials (these trajectories are 
investigated in \cite{BD-inv-perturbations}). A totally classical 
analysis therefore enables us to write the orbits of the fields in 
terms of the initial conditions ($\sigma_0$,$\Phi_0$), 
and the motion is restricted to one degree of freedom 
on the plane ($\sigma$,$\Phi$). In addition to the classical motion, 
quantum diffusion is responsible for "jumps" of the 
fields between classical trajectories, and therefore a large 
number of classical trajectories are accessible after 
a certain period of evolution, as is predicted in the 
self-reproducing universe model.\\
\\
In the case of a powerlaw potential $V(\sigma)=\lambda/(2n)\,\sigma^{2n}$ 
the curve (\ref{conservation}) becomes
\be \label{curve} 
\Phi=\biggl(\Phi_0+{1\over 4n\omega}\,\sigma_0^2\biggr)-{1\over 4n\omega}
\,\sigma^2,
\ee
as is shown in Fig.~1, with the corresponding BoI and EoI curves 
(\ref{BoI})(\ref{EoI}). The EoI curve is always a parabola, 
whereas 
the BoI curve is a straight line for $n=1$, a parabola for 
$n=2$. For $n>2$ both curves intersect at a certain value $\Phi_{\rm max}$, 
given by 
\be \label{phimax} 
\Phi_{\rm max}={1\over4 n^2}\,\biggl({3\omega-2\over\omega}
\biggr)^{n/(n-2)}\biggl({32\pi^2\over \lambda n^3}\biggr)^{1/(n-2)}. 
\ee
Therefore the region between the curves BoI and EoI on the 
($\sigma$,$\Phi$) 
where inflation takes place is bounded in the case of $n>2$ 
(region enclosed between the thick solid and dotted lines in Fig.~1). 
Inflation will set on if the initial 
BD field is within the range $0<\Phi_0<\Phi_{\rm max}$ on the 
curve BoI and the period of inflation decreases steadily until it 
vanishes at $\Phi_0=\Phi_{\rm max}$. The alternative case 
of $n\leq 2$ gives rise to the so-called `run-away' solutions 
that we discuss in the following section. \\
\\
The ratio of energy densities of the fields $\sigma$,$\Phi$ is 
\be \label{energies}
{\rho_\Phi\over\rho_\sigma}\sim {\Phi\over 3\omega^2},
\ee
and whereas the energy density of $\sigma$ is dominated by the 
potential energy, $\Phi$ is entirely driven by its kinetic 
energy. From (\ref{energies}) we see that at large $\Phi$ the 
energy density of the field $\Phi$ overcomes that of inflation. 

\subsection{`Run-away' solutions}

These solutions occur in the case of $\Phi_{\rm max}\to\infty$ 
($n\leq 2$ in the case of a powerlaw potential). Even though 
the classical trajectory (\ref{curve}) is a parabola for an 
arbitrary large initial BD field $\Phi_0$, quantum diffusion 
drives the fields to larger and larger values and, for as long as 
inflation takes place, a tendency towards greater values 
of $\Phi$ is enhanced \cite{BD-inv-perturbations}. Quantum 
jumps that take the fields from the initial classical trajectory 
to other classical trajectories corresponding to larger values 
of the initial fields are favoured. This process goes on 
indefinitely and the period of inflation is prolonged for those 
regions where jumps to larger values of the fields take place. 
Therefore, the largest physical volumes are occupied by values 
of the fields that grow without limit, where the period of inflation 
is indefinitely long. 
These so-called {\it run-away solutions} do not permit us 
to make predictions on the most typical values of the fields. 
The volume of the universe is almost in its entirety 
occupied by regions of arbitrarily large $M_P$. These solutions 
are only compatible with the observed universe with negligible 
probability and are therefore ruled out.\\
\\
For $n>2$, as seen in Fig.~1, 
the BoI and EoI boundaries (\ref{BoI})(\ref{EoI}) 
intersect at a value $\Phi=\Phi_{\rm max}$, and it is   
that value of $\Phi$ that occupies the largest fraction of the 
total physical volume. These theories make a definite 
prediction of the likeliest Planck mass at the end of inflation 
\be \label{pm}
M_{P*}^2=16\pi\,\Phi_{\rm max},
\ee 
where $\Phi_{\rm max}$ is given by (\ref{phimax}). 
In the limit of large $\omega$, $\lambda$ thus satisfies the following 
relation 
\be  \label{lambda-mag}
\lambda= 2n\, \biggl({12\pi\over n^2}\biggr)^n
\biggl({1\over M_{P*}^2}\biggr)^{n-2}.
\ee
Therefore, if a $n>2$ powerlaw potential is the right theory 
and the Planck mass (\ref{pm}) is given by its value in our 
region of the universe, $M_{P*}^2\sim 10^{19}$ GeV,  
under the assumption that our immediate neighbourhood is a 
{\it typical} 
region, $\lambda$ results in the following order of magnitude  
\be \label{lambda}
\lambda\sim 10^{-17}
\ee
for $n=3$. This is effectively an upper limit for the 
order of magnitude 
of $\lambda$ predicted by (\ref{lambda-mag}), as it decreases sharply 
for greater values of $n$. We can also conclude from (\ref{lambda-mag})  
that larger values of $\lambda$ can only be realistic if the 
typical value of $M_{P*}^2$ is several orders of magnitude 
smaller than the value measured in our region of the universe. 

\subsection{Stationary universe}

The extended inflation theory describes a self-reproducing 
universe where the values of the fields $(\sigma,\Phi)$ are 
described in terms of probability distributions. In this section, 
we will summarize some results (see e.g. \cite{BD-inv-perturbations,
BD2,BD-perturbations}). The comoving probability $\calP_c(\sigma,\Phi,t)$ 
satisfies the convervation equation \cite{inflation-general,stochastic}
\be \label{prob-conservation}
\partt\calP_c= -\parts J_{\sigma}-\partp J_{\Phi},
\ee
where the probability current $\b J\equiv (J_{\sigma},J_{\Phi})$ is 
given by, in the slow-roll regime where the effect of 
quantum diffusion can be neglected, 
\be \label{jsigma}
J_{\sigma}\approx -{M^2_P(\Phi)\over 4\pi}\,H^{\alpha-1} 
\parts H\,\calP_c,
\ee
\be \label{jphi}
J_{\Phi}\approx -{M^2_P(\Phi)\over 2\pi}\, H^{\alpha-1}
\partp H\,\calP_c,
\ee
The index $\alpha$ denotes the choice of time parametrization. The 
synchronous gauge is recovered in the case of $\alpha=0$, which corresponds 
to $t= \log a$, and $\alpha=1$ corresponds to proper or cosmic time, 
$t=\tau$. Henceforth we will adopt the synchronous gauge. As it is 
apparent from the results discussed in \cite{BD2}, the probability 
distributions are very sensitive to the choice of time variable. 
Nonetheless they are helpful in giving a qualitative picture of 
the dynamics of the fields and whether they reach asymptotic 
values or grow indefinitely. Also we shall see 
that the physical probabilities (or equivalently, 
the fraction of the physical volume occupied by a homogeneous 
hypersurface ($\sigma$,$\Phi$)=const) can be computed 
in a way that is insensitive to the choice of time parameter. This 
procedure was first suggested in \cite{method} for an inflation-only 
scenario, and implemented in the case of extended inflation in 
\cite{BD-inv-perturbations}. 
In order to transform (\ref{prob-conservation}) 
into an eigenvalue equation, we 
consider the following expansion: 
\be \label{pc}
\calP_c(\sigma,\Phi,t)=\sum_{n=1}^\infty \psi_n(\sigma,\Phi)\,
\e^{-\gamma_nt},
\ee
where $\gamma_1<\gamma_2<\gamma_3<...$, and thus in the 
limit $t\to\infty$ the dominant contribution is that of $n=1$, 
$\calP_c\sim \psi_1\,\e^{-\gamma_1t}$. The value of $\gamma_1$ 
depends on the form of the potential, and is determined 
by the boundary condition that establishes the conservation of 
probability flux along the EoI boundary. For a powerlaw 
potential typically we have $7.8\lsim\gamma_1\lsim 8.3$ for 
$10^{-18}\lsim\lambda\lsim 10^{-15}$ (we use these small values 
of $\lambda$ in view of (\ref{lambda-mag})).  
By substituting (\ref{jsigma})(\ref{jphi}) into (\ref{prob-conservation}), 
it is easy to calculate the following asympotic solution: 
\be \label{comoving} 
\calP_c(\sigma,\Phi,t)\sim C_0\,\Phi^{{\omega\over 2}\gamma_1-1}\,
\biggl({V\over V'}\biggr)\,\e^{-\gamma_1 t}, 
\ee
where $C_0$ is a normalization constant. 
The comoving probability 
(\ref{comoving}) gives us an idea of the likelihood of certain 
configurations in different regions of the universe. However, 
for those regions in the $(\sigma,\Phi)$ plane that lie between 
the BoI and EoI curves (\ref{BoI})(\ref{EoI}), 
the physical volumes are many orders of magnitude greater than 
those regions that have undergone thermalization at the same 
comoving scale. Therefore, in order to address the question of 
relative likelihood of certain physical quantities (such as 
the value of the Planck mass), we need to look into the 
actual physical volumes of homogeneous hypersurfaces. 
By virtue of the principle of stationarity, 
the fraction of the volume of a given hypersurface with respect 
to the total volume reaches an asymptotic value, and it is therefore 
possible to examine the question of how typical a quantity is 
in terms of volume ratios.\\
\\                                                     
There are two equivalent ways to tackle this problem. The first 
one is to write (\ref{prob-conservation}) in physical coordinates, 
for a physical probability distribution $\calP_p$ 
(rather than the comoving $\calP_c$), by adding an extra 
term $3H\calP_p$ on the RHS of (\ref{prob-conservation}). The 
presence of this extra term, that accounts for the background 
expansion, complicates the problem and the eigenvalue equation is 
no longer soluble analytically.\footnote{A gauge-invariant 
approach for calculating the spectrum of density fluctuations was 
pursued in \cite{BD-perturbations} along these lines, without 
an explicit derivation of $\calP_p$.} 
An second procedure, that we will use here, 
is to compute the physical volumes of 
the hypersurfaces ($\sigma$,$\Phi$)=const. Given a suitable 
normalization, the volumes of homogeneous hypersurfaces  
can be expressed in terms of the fraction of the total volume 
or the volume of the thermalized regions 
\cite{method,method-challenge}. As it is shown in the  
Appendix, the ratio of the physical volume $\v(\sigma,\Phi)$ 
of an arbitrary hypersurface with respect to the thermalized 
volume $\v_*$ for a powerlaw potential is given by 
\be \label{ratio}
r(\sigma,\Phi)
={\v(\sigma,\Phi)\over \v_*}\sim \Phi^{{\omega\over 2}\gamma_1}
\,\biggl(1+{\sigma^2\over 4n^2\,\omega^2}\biggr)^{1/2}.
\ee
The likelihood of a point lying on the hypersurface 
($\sigma$,$\Phi$)=const is proportional to its volume, 
so $r$ gives a measure of this likelihood. Comparing (\ref{ratio}) 
and (\ref{comoving}), we notice that in both cases the 
same tendency is preserved, that of the field $\sigma$ 
rolling down towards the minima of $V(\sigma)$, whereas 
the BD field $\Phi$ tends to increase towards the largest 
values possible. 

\section{Variable $\omega$: hyperextended inflation}

In this section we generalize the results of \S 2 for a dynamical 
$\omega$. This theory was suggested by \cite{hyperI,hyperII}, motivated by 
observational discrepancies of the extended inflation model 
pointed out in \cite{constraints}. The CMB data analysis of 
\cite{constraints} requires $\omega\leq 25$, as opposed to the 
constraint $\omega>500$ of \cite{time-delay}, as was discussed 
in the Introduction. The hyperextended inflation action is 
again given by (\ref{action}), where $\omega$ is dependent on 
$\Phi$. 
The resulting equations are 
\be \label{fieldI}
R_{\mu\nu}-{1\over 2}R\,g_{\mu\nu}= {\omega(\Phi)\over\Phi^2}
\,\biggl[\partial_{\mu}\Phi\,\partial_{\nu}\Phi-{1\over 2}
g_{\mu\nu}\,(\partial\Phi)^2\biggr]
+{1\over\Phi}
\,(\nabla_{\mu}\nabla_{\nu}\Phi-g_{\mu\nu}\,\Box\Phi)
+{8\pi\over\Phi}\,T_{\mu\nu},
\ee
\be \label{fieldII}
\Box\Phi={1\over 2\omega(\Phi)+3}\,\biggl[8\pi T_{\mu}^{\mu}
-\omega'(\Phi)(\partial\Phi)^2\biggr],
\ee
\be \label{fieldIII}
\Box\sigma= -V'(\sigma),
\ee
where the energy-momentum tensor of the matter sector 
$T_{\mu\nu}$ is given by
\be
T_{\mu\nu}={1\over 16\pi}\,\biggl[\partial_{\mu}\sigma
\,\partial_{\nu}\sigma-{1\over 2}g_{\mu\nu}\,(\partial\sigma)^2
+ g_{\mu\nu}\,V(\sigma)\biggr].
\ee
We examine the solutions in a FRW metric:
\be
\d s^2= \d t^2- a^2(t)\biggl[{\d r^2\over 1-kr^2} +r^2\,
(\d\theta^2+\sin^2\theta\,\d\phi^2)\biggl],
\ee
where $k=0,\pm 1$ and $a(t)$ is the scale factor of the universe.  
Then the field equations (\ref{fieldI})-(\ref{fieldIII}) 
become
\be \label{FRWI}
H^2 +{k\over a^2}+ H\,{\dot\Phi\over\Phi}= {1\over 6}\,\omega
\,{\dot\Phi^2\over\Phi^2}+{1\over 6\Phi}\,\biggl[V(\sigma)
+{1\over 2}\dot\sigma^2\biggr],
\ee
\be  \label{FRWII}
\dot H+ 3H^2+2{k\over a^2}+ {5\over 2}H\,{\dot\Phi\over\Phi}=  
-{1\over 2}\,{\ddot\Phi\over\Phi}+{1\over 2\Phi}\,V(\Phi),
\ee
\be \label{FRWIII}
\ddot\Phi+ 3H\,\dot\Phi = {1\over 2\omega+3}\,
\biggl[2V(\sigma) -{1\over 2}\,\dot\sigma^2-\omega'\,\dot\Phi^2\biggr],
\ee
\be \label{FRWIV}
\ddot\sigma + 3H\,\dot\sigma = -V'(\sigma),
\ee
where, as is customary $H\equiv {\dot a/a}$. In the slow-roll 
approximation, (\ref{FRWI})-(\ref{FRWIV}) become
\be \label{SRR1}
{\dot\Phi\over\Phi}= 2\Sigma\,{H\over\omega},
\ee
\be \label{SRR2} 
\dot\sigma= -2\Phi\,{V'\over V}\,H,
\ee
\be \label{SRR3} 
H^2= {V(\sigma)\over 6\Phi},
\ee
where $\Sigma\equiv (1-\Phi\omega')$. 
These equations are essentially identical to (\ref{SR1})-(\ref{SR3}), 
with the exception of the $\Sigma$ factor in (\ref{SRR1}). 
For negligible variations of $\omega$, such that $\omega'\ll\Phi^{-1}$, 
both models are naturally the same, i.e. $\Sigma=1$. The departure 
of $\Sigma$ from unity is the characteristic of the hyperextended model. 
The EoI boundary is given by 
\be
{1\over 2}\,\dot\sigma^2+{\omega\over\Sigma^2}\,
{\dot\Phi^2\over\Phi}= V(\sigma),
\ee
that yields, with the aid of (\ref{SRR1})-(\ref{SRR3}), 
\be \label{hyper-EoI}
\Phi_*=\biggl({3\omega_*-2\over\omega_*}\biggr)
\biggl({V\over V'}\biggl)_*^2, 
\ee
where as usual $*$ denotes the values of the quantities at the end 
of inflation. It is easy to see that (\ref{hyper-EoI}) is in 
fact formally equivalent to (\ref{EoI}), only that $\omega_*$ 
is dependent on $\Phi_*$, so (\ref{EoI}) is solved for $\Phi_*$ once 
the functional dependence of $\omega$ is determined.  
The BoI boundary is given by (\ref{BoI}) as in extended inflation. 
From (\ref{SRR1})(\ref{SRR2}) the following conservation 
law follows
\be \label{conservation2}
{\d\over\d t}\biggl(\int\! {\omega\over\Sigma}\,\d\Phi
+\int\!{V\over V'}\,\d\sigma\biggr)=0,
\ee
which is a generalization of (\ref{conservation}). The form of 
the orbits given by (\ref{conservation2}) will strongly depend 
on the model for $\omega(\Phi)$ that we use. Qualitatively, we 
see from (\ref{conservation2}) that the classical orbits  
depart more from the extended inflation ones (\ref{conservation}) 
for smaller values of $\Phi$ (corresponding to early stages 
of inflation) than larger ones (close to EoI boundary).  
In principle the 
hyperextended inflation action does not contain sufficient information 
to determine the variations of $\omega$, and we need to make 
an Ansatz for its functional dependence. In the following sections 
we will investigate this problem.

\subsection{Principle of least coupling}

In string theory the BD field $\Phi$ appears naturally in the 
form of the dilaton $\Psi$, a scalar field associated to the graviton 
tensor field. The effective action for the graviton-dilaton-inflation 
sector is given by \cite{string}
\be \label{s-action} 
{\cal S} = \int\!\d^4 x\,\sqrt{-g}\,\e^{-2\Psi}\,
\biggl[B_g(\Psi)\, R +B_{\Psi}(\Psi)\, 4(\partial\Psi)^2 
-{1\over 2}\,B_{\xi}(\Psi)\,(\partial\sigma)^2
-V(\Psi,\sigma)\biggr],
\ee
where it is assumed that the string tension $\alpha=1$; 
$B_i(\Psi)$ are coupling functions. 
The matter couplings $B_i(\Psi)$ of the dilaton, which play a r\^{o}le 
equivalent to the BD coupling $\omega(\Phi)$, are responsible 
for deviations from GR. Little is known of the 
general form of $B_i(\Psi)$, which depend on the details of the 
perturbative higher-loop corrections of the scalar-tensor interactions. 
However, a simple assumption 
of universality of these coupling functions, such that 
$B_i(\Psi)=B(\Psi)$ leads to a simple and interesting model. 
The cosmological evolution of the graviton-dilaton-matter 
system under this assumption drives the dilaton to a 
massless regime (as is shown in \cite{least-coupling}) 
and thus it decouples from matter asymptotically. This 
is also known as the {\it principle of least coupling}, that we 
will assume in this paper by applying it to $\omega(\Phi)$. 
The coupling function $B(\Psi)$ is therefore given as a Taylor 
expansion in the string coupling $g^2_s\equiv \e^{2\Psi}$ 
\cite{least-coupling}:
\be
B(\Psi)= b_0+b_1\,\e^{2\Psi}+b_2\,\e^{4\Psi}+\dots .
\ee
These expressions translate into the BD formalism in the 
following way. The dilaton $\Psi$ relates to the BD field $\Phi$ 
via 
\be
\Phi\equiv\exp (-2\Psi),
\ee
and thus $g_s^2=\Phi^{-1}$. The BD coupling $\omega(\Phi)$ is 
given then by 
\be \label{w-taylor} 
\omega(\Phi)= \eta_0+{\eta_1\over\Phi}+{\eta_2\over\Phi^2}
+{\eta_3\over\Phi^3}+\dots ,
\ee
where $\eta_0$ is given by the low-energy value predicted 
by string theory and the higher-order coefficients $\eta_i$ are 
determined by string-loop corrections. Although $\eta_0$ is 
strongly dependent on the mechanisms of compactification 
and supersymmetry breaking, it is widely accepted that 
$\eta_0= -1$ in four dimensions. We must rely on observational 
constraints or string-loop calculations to estimate the remaining 
$\eta_i$.\\
\\
In a cosmological inflationary model, we have seen 
in \S 2 that the course of inflation leads the BD field to grow, 
in some cases without limit. Thus $g_S^2$ is a good perturbation 
parameter and it is apparent from (\ref{w-taylor}) 
that in a hypothetical asymptotic case of $\Phi\to\infty$, the 
most probable value of $\omega$ is $-1$ and any other 
value takes place with negligible probability. A large 
positive value of $\omega$ can be probable only if the BD field 
is bounded above, i.e. $\Phi_{\rm max}< \infty$, and provided 
that the contribution of the terms $\eta_n/\Phi_{\rm max}^n$ in 
(\ref{w-taylor}) is not negligible with respect to unity. 
We have already seen in \S 2 that in extended 
inflation it is also the case that only models with finite 
$\Phi_{\rm max}$ are realistic from the astrophysical point of 
view. \\
\\
Let us investigate the following toy model. 
We assume that a large value of $\omega$ is 
pertinent so that it is consistent with the observational bound 
$\omega\gg 500$, and the inflation potential such that $\Phi$ 
is bounded above, 
e.g. a powerlaw potential $V(\sigma)=\lambda/(2n)\,
\sigma^{2n}$ with $n>2$. The coupling $\omega(\Phi)$ is 
truncated for simplicity to the one-parameter Ansatz 
\be \label{toyI}
\omega(\Phi)= -1+{\eta\over\Phi},
\ee
and $\eta>0$. Inflation occurs for values of the BD 
field that grow from an initial value $\Phi_0$ to a maximum value 
$\Phi_{\rm max}$ that is given by\footnote{We use (\ref{phimax}) 
under the approximation $(3\omega_*-2)/\omega_*\to 3$ for large 
$\omega$. Therefore the EoI boundary is 
$\Phi_*\approx 3\sigma_*^2/4n^2$}
\be \label{hyp-phimax}
\Phi_{\rm max} = {3^{n/(n-2)}\over 4n^2}\,\biggl({32\pi^2\over\lambda n^3}
\biggr)^{1/(n-2)}.
\ee 
Most regions are occupied by $\Phi_{\rm max}$,  
as is shown in \S 2.1. In these regions $\omega_{\rm typical}\gg 1$, and 
thus $\eta\approx \Phi_{\rm max}\,\omega_{\rm typical}$. Therefore  
\be  \label{omega-exp}
\omega(\Phi)\approx \omega_{\rm typical}\,{\Phi_{\rm max}
\over\Phi}.
\ee
This toy model is schematically illustrated in Fig.~2. 
The value of $\omega$ 
decreases monotonically during the course of inflation from 
a given arbitrary value $\omega_0$ to its EoI inflation value, 
$\omega(\Phi_*)$, the most likely of which is 
$\omega_{\rm typical}=\omega(\Phi_{\rm max})$. 
On the other hand, if $\omega_0\approx -1$, such that 
$\eta\ll\Phi_0$, then $\omega\approx -1$ throughout. \\
\\
From the conservation law (\ref{conservation2}) follows that 
the trajectories of the fields are given by 
\be \label{worbits}
2\eta\,\log(\eta+\Phi)-\Phi+{1\over4 n}\,\sigma^2 ={\rm const}.  
\ee 
These trajectories reduce to the parabolical orbits 
described by (\ref{conservation}) in the approximation 
$O(\Phi^2/\eta^2)$, whereas the following order of the expansion 
of the logarithmic term in (\ref{worbits}) leads to 
\be \label{curve2}
\Phi-{1\over\eta}\,\Phi^2 +{1\over 4n}\,\sigma^2={\rm const}. 
\ee
It is easy to see that, in comparison to (\ref{curve}), 
for sufficiently large $\eta$, the trajectories (\ref{curve2}) 
lead to larger final values of 
$\Phi$ relative to the parabola (\ref{curve}) for similar initial 
conditions. This effect can only be reproduced in (\ref{curve}) by means of 
quantum fluctuations that take the fields from one parabola to an outer 
one where the fields are larger. The approximation (\ref{curve2}) 
enables us to constrain $\eta$ (and therefore 
$\omega_{\rm typical}$) in this simple model in terms of the 
initial conditions $(\sigma_0,\Phi_0)$, given that $\Phi_{\rm max}$ 
is fixed by the choice of potential. In qualitative terms, it 
can be also said that due to the growth of $\Phi$ during the 
course of inflation, $\omega$ is always greater at the initial time 
(in the model described by (\ref{toyI})) than at the end of inflation.

\subsection{Probability distributions and volume ratios}

We now compute the probability distributions and volume ratios 
of homogeneous hypersurfaces, in order to generalize the results of \S 2 
for variable $\omega$. In the following subsection we will apply 
these results to the simple one-parameter Ansatz that we have 
briefly discussed above. From (\ref{SRR1})-(\ref{SRR3}) it is 
easy to see that the probability current ($J_{\sigma}$,$J_{\Phi}$) is 
given by 
\be \label{hypI}
J_{\sigma}= -2\Phi\,\biggl({V'\over V}\biggr)\,\calP_c,
\ee
\be \label{hypII}
J_{\Phi}= 2\Phi\,\biggl({\Sigma\over\omega}\biggr)\,\calP_c.
\ee
The continuity equation (\ref{prob-conservation}) then yields, in 
the limit $\calP_c\sim \Psi_1\,\e^{-\gamma_1\,t}$,
\be
\calP_c(\sigma,\Phi,t)= C_0\, \biggl({V\over V'}\biggr)\,
\Phi^{-1}\,\biggl({\omega\over\Sigma}\biggr)\,
\exp\biggl({\gamma_1\over 2}\int\!{\omega\over\Phi
\Sigma}\,\d\Phi -\gamma_1 t\biggr). 
\ee
Furthermore, (\ref{hypI})(\ref{hypII}) enable us to 
compute the regularized volumes of thermalized regions and 
homogeneous hypersurfaces via (\ref{volume1})(\ref{volume2}). 
Once again we focus on the particular case of a powerlaw potential.  
It is easy to show that the thermalized volume is, to a 
good approximation, insensitive to variations of $\omega$, and therefore 
it is given by the extended inflation result (\ref{volume22}), where 
$\omega$ is evaluated at $\omega(\Phi_{\rm max})$. 
The volume of a homogeneous hypersurface is on the other 
hand  
\be \label{hyp-volume}
\v(\sigma,\Phi)_{\rm regularized} = 2\v_0\;\biggl|
{\e^{(3-\gamma_1)\,t_c}-1\over 3-\gamma_1}\biggr| \,
\biggl({\omega\over\Sigma}\biggr)\,
\biggl(1+{\sigma^2\over 4n^2\omega^2}\,\Sigma^2\biggr)^{1/2}
\,\exp\biggl({\gamma_1\over 2}\int\!
{\omega\over\Phi\,\Sigma}\,\d\Phi -\gamma_1 t\biggr),
\ee
where $t_c$ is the cut-off time as is explained in the 
Appendix. The volume 
ratio $r$ of a homogeneous hypersurface with respect to the thermalized 
volumes is then
\be \label{hyp-ratio}
r(\sigma,\Phi)= {\gamma_1\over 2}\,\Phi_{\rm max}^{-{\omega\over 2}\,
\gamma_1-1}\,\omega(\Phi_{\rm max})\,
\biggl({\omega\over\Sigma}\biggr)\,
\exp\biggl({\gamma_1\over 2}\int\! {\omega\over\Phi\,\Sigma}
\,\d\Phi\biggr)
\biggl(1+ {\sigma^2\over 4n^2\omega^2}\,
\Sigma^2\biggr)^{1/2}.  
\ee
It is straightforward also from (\ref{hyp-ratio}) to compute the 
relative likelihoods of ($\sigma$,$\Phi$) and ($\tilde\sigma$,
$\tilde\Phi$) by working out the ratio $\tilde{r}/r$. For two 
homogeneous hypersurfaces that are only differentially separated, 
this ratio becomes
\be
{\tilde r\over r}= 1+{\sigma\over 4n^2}\,\delta\sigma 
-{2\over\Phi}\,\delta\Phi.
\ee
By comparing (\ref{hyp-ratio}) with (\ref{ratio}), we observe that 
the multiple presence of the $\Sigma(\Phi)$ factor in (\ref{hyp-ratio}) 
can potentially yield a very different result, in comparison to 
extended inflation, depending on $\omega(\Phi)$. 
This factor is, from (\ref{w-taylor}),  
\be \label{hyp-factor}  
\Sigma(\Phi)= 1+{\eta_1\over\Phi}+2\,{\eta_2\over\Phi^2} 
+3\,{\eta_3\over\Phi^3}+\dots ,
\ee
and therefore one recovers extended inflation, $\Sigma\approx 1$,  
in the limit 
$\Phi_{\rm max}\to\infty$ and also, as $\Phi$ increases throughout 
the course of inflation, the predictions of (\ref{hyp-ratio})  
differ less from those of (\ref{ratio}) the closer we get to the EoI 
boundary, and they differ most at the early stages of inflation. 
In the most general case (\ref{w-taylor}), the ratio 
(\ref{hyp-factor}) requires a numerical resolution, but we can see 
that the result is roughly of the form
\be
r(\sigma,\Phi)\sim \Pi_i\;(f_i-\Phi)^{e_i}\,\biggl(1
+{\sigma^2\over4n^2\omega^2}\,\Sigma^2\biggr),
\ee
where $f_i$ are the poles of the ratio $\omega/\Sigma$ and $e_i$ are 
real numbers. The following toy model illustrates the simplest 
non-trivial scenario. 

\subsection{Toy model}

Let us consider the simplest parametrization (\ref{toyI}) that we have 
discussed at the end of \S 3.1 in a model where 
$\omega$ is in most regions sufficiently large, so that $\eta\gg\Phi$, 
and therefore $\omega\approx \eta/\Phi$. Thus, $\Sigma\approx\omega$. 
The EoI boundary is then 
given by $\Phi\approx 3\sigma^2/(2n)^2$ and it follows that 
\be \label{toy-prob}
\calP_c(\sigma,\omega,t)= C_0\,\biggl({\sigma\over 2n\eta}\biggr)
\,\omega\,(\omega+1)^{{\gamma_1\over 2}+1}(\omega+2)^{-\gamma_1-1}
\,\e^{-\gamma_1\,t}
\sim \sigma\,\omega^{1-{\gamma_1\over 2}}\,\e^{-\gamma_1\,t},
\ee
\\
and the corresponding volume ratio 
\be \label{toy-ratioI}
r(\sigma,\omega)\sim \biggl({\omega^2\over\omega +2}\biggr)\,
\biggl[{\omega +1\over (\omega+2)^2}\biggr]^{\gamma_1/2}\,
\biggl[1+ {\sigma^2\over 4n^2}\,\biggl({\omega+2\over\omega}\biggr)^2
\biggr]^{1/2}\sim \omega^{1-{\gamma_1\over 2}}\,
\biggl(1+{\sigma^2\over 4n^2}\biggr)^{1/2}.
\ee
 For a typical value $\gamma_1\sim 8$, 
(\ref{toy-prob})(\ref{toy-ratioI}) predict $r\sim \omega^{-3}$, 
i.e. smaller values of $\omega$ are likelier than larger ones and,  
as indeed according to the Ansatz (\ref{toyI}) $\omega$ decreases 
during the course of inflation, the most typical value within 
this model is $\omega_*=\eta/\Phi_{\rm max}$, where $\Phi_{\rm max}$ 
is as before given by (\ref{hyp-phimax}). A tendency towards 
greater $\omega$ requires a fine-tuning of parameters to 
orchestrate $\gamma_1<2$, at any rate in contradiction with 
the inequality $\gamma_1>3$ that must be satisfied for inflation 
to be eternal. Following (\ref{toy-ratioI}) therefore most regions in the 
universe are occupied by $\omega(\Phi_{\rm max})$, that prevails 
as the most typical value of $\omega$ after inflation. In conclusion, 
$\omega(\Phi)$ is determined via (\ref{omega-exp}) 
by the astrophysical determination of $\omega_{\rm typical}$ and a 
choice of potential that yields $\Phi_{\rm max}$. \\
\\
Another toy model to the next order in (\ref{w-taylor}) is 
\be \label{toyII}
\omega= -1+{\eta_1\over\Phi}+{\eta_2\over\Phi^2},
\ee
where again we assume that $\omega$ is large compared to unity. 
(\ref{toyII}) can be inverted, i.e. 
\be
\Phi\approx {\eta_1\over 2\omega}\,\biggl[1+\biggl(1+4\,
{\eta_2\over\eta_1^2}\,\omega\biggr)^{1/2}\biggr],
\ee
or if $\eta_1$,$\eta_2$ are of comparable magnitude, then  
we have $\Phi\approx (\eta_2/\omega)^{1/2}$ and
$\Sigma\approx 2\omega$. Therefore  
\be \label{toy-probII}
\calP_c(\sigma,\omega,t)\approx C_0\,\biggl({\sigma\over 4n}\biggr)
\,\biggl({\omega\over\eta_2}\biggr)^{1/2-\gamma_1/8}\,\e^{-\gamma_1\,t},
\ee
and 
\be \label{toy-ratioII}
r(\sigma,\omega)\approx \omega^{1-\gamma_1/8}\,\biggl(1+
{\sigma^2\over 4n^2}\biggr)^{1/2}.
\ee
In (\ref{toy-ratioII}) we find the same tendency as in (\ref{toy-ratioI}) 
towards smaller values of $\omega$. This tendency is less manifest 
in (\ref{toy-ratioII}) and for a conservative value of $\gamma_1=8$  
(\ref{toy-ratioII}) appears to be independent of $\omega$, due to 
the crude approximation $\eta_1\sim\eta_2$ involved in the derivation 
of (\ref{toy-probII})(\ref{toy-ratioII}). A more detailed numerical 
derivation of (\ref{toy-ratioII}) shows that for arbitrary $\eta_1$,$\eta_2$, 
in fact one obtains $r\sim \omega^{1-\gamma_1/\alpha}$, with 
$3.4<\alpha<3.9$, which is compatible with the conclusions of 
the simplest toy model (\ref{toyI}). 

\subsection{Asymptotic regime}

In the extreme case where $\Phi_0\sim 0$ and $\Phi_{\rm max}\to\infty$, 
it is easy to see that any given Ansatz of the type
\be \label{w-truncated}
\omega(\Phi)= -1+{\eta_1\over\Phi}+{\eta_2\over\Phi^2}+\dots+
{\eta_M\over\Phi^M},
\ee
i.e. a truncated version of (\ref{w-taylor}), is dominated 
by the lowest and highest orders in the asymptotic regimes:
\be
\omega(\Phi_0)\approx {\eta_M\over\Phi_0^M}
\ee
and 
\be \label{phi-astro}
\omega(\Phi_{\rm max})\approx -1+{\eta_1\over\Phi_{\rm max}}.
\ee
If it is correct to assume that at the initial time the homogeneous 
bubble that undergoes inflation is in a string theory ground state, 
then $\omega(\Phi_0)\approx-1$ and $\eta_M\approx -\Phi_0^M$. 
From the astrophysical point of view, (\ref{phi-astro}) represents  
the asymptotic behaviour near the EoI boundary, and is of interest 
as most regions of the universe are occupied by $\Phi=\Phi_{\rm max}$. 
Therefore, in the asymptotic regime any model of the type (\ref{w-truncated}) 
is reduced to the toy model (\ref{toyI}) that we have discussed in the 
previous section. \\
\\
The relative values of the coefficients $\eta_i$ cannot be determined 
directly from the hyperextended inflation action and they are 
fixed by higher-loop estimates in string theory. If $\eta_2/\eta_1\gg 1$, 
then the relative magnitude of $\eta_2$ with respect to $\Phi_{\rm max}$ 
determines whether (\ref{phi-astro}) requires a higher-order correction 
by adding the term $\eta_2/\Phi_{\rm max}^2$ on the RHS. So far our 
calculations are based on the assumption that the $\eta_i$ are 
well-behaved and therefore (\ref{phi-astro}) can be considered 
a good approximation given these provisos.

\subsection{Spectrum of fluctuations}

An ensemble of observers located on a homogeneous hypersurface 
($\sigma$,$\Phi$)=const observes quantum jumps of the fields due 
to the stochastic nature of inflation. On the one hand, there is 
the contribution of quantum fluctuations stretched beyond the horizon 
distance, that is effectively a stochastic force that acts on the classical 
solutions (\ref{SRR1})-(\ref{SRR3}), such that 
\be \label{stochI}
\dot\sigma = -2\Phi\,\biggl({V'\over V}\biggr)
\,H +{H^{3/2}\over 2\pi}\,\zeta(t),
\ee
\be \label{stochII}
\dot\Phi=2\,\biggl({\Sigma\over\omega}\biggr)\,H\Phi
+{H^{3/2}\over 2\pi}\,\xi(t),
\ee
where $\zeta$,$\xi$ follow a Gaussian distribution, and 
$\langle\zeta(t_1)\zeta(t_2)\rangle=\langle\xi(t_1)\xi(t_2)\rangle
=\delta(t_1-t_2)$ and $\langle\zeta(t_1)\xi(t_2)\rangle=0$. The second terms 
on the RHS of (\ref{stochI})(\ref{stochII}) are random fluctuations 
of the fields that are superimposed on
the slow-roll [classical] solutions over distances greater than 
$H^{-1}$ (first terms on the RHS of 
(\ref{stochI})(\ref{stochII})). This is the so-called "coarse-grained" 
description of the fields. On the other 
hand, quantum jumps that take the fields to greater values are  
likelier than those that take them to smaller ones, because the physical 
volume occupied by the former is far greater. 
The volume ratio's dependence on the fields is 
rather steep in most cases, so fluctuations that end up in 
hypersurfaces of greater values are enhanced. This enhancement 
factor is approximately $\sim \v_B/\v_A$, where $\v_A$ is the 
volume of the hypersurface where the observers are located and 
$\v_B$ the volume of the hypersurface where the fields end up as 
the result of a fluctuation ($\delta\sigma$,$\delta\Phi$). 
As a consequence of the combination of this factor and the stochastic 
nature of inflation, the typical quantum jumps ($\delta\sigma$,$\delta\Phi$) 
observed by the average observer follow the distribution
\be  \label{real-prob}
\d\calP(\delta\sigma,\delta\Phi)\sim 
{\v(\sigma+\delta\sigma,\Phi+\delta\Phi)\over \v(\sigma,\Phi)}\,
\d\calP_0(\delta\sigma,\delta\Phi),
\ee
where $\d\calP_0$ is the Gaussian probability distribution of 
the stochastic field fluctuations, 
\be  \label{gauss-prob}
\d\calP_0(\delta\sigma,\delta\Phi)={1\over (2\pi\Delta)^{1/2}}
\,\exp\biggl[-{(\delta\sigma)^2+(\delta\Phi)^2\over 2\Delta^2}\biggr]\,
\d\delta\sigma\,\d\delta\Phi,
\ee
and the variance of the fields 
$\Delta=\langle\delta\sigma\rangle=\langle\delta\Phi\rangle
=H/(2\pi)$. The distribution (\ref{real-prob}) is clearly non-Gaussian, 
as it is the product of a non-Gaussian distribution (the volume ratios) 
and a Gaussian one (the stochastic motion of the fields).\\
\\
The stationary values of (\ref{real-prob}) yield the 
expectation values of the quantum jumps, $\langle\sigma\rangle$ 
and $\langle\Phi\rangle$, out of which we compute the spectrum 
of density fluctuations. The volume ratio 
$\v(\sigma+\delta\sigma,\Phi+\delta\Phi)/\v(\sigma,\Phi)$ has 
in general a complicated form, as can be seen from (\ref{hyp-volume}) 
and we simplify the calculation by examining the toy model 
(\ref{toyI}) for a powerlaw potential. As we have seen in \S 3.3-4, 
(\ref{toyI}) is a good approximation near the EoI boundary, which 
is the regime of interest for the spectrum of fluctuations. 
In this case, 
the volume ratio of the two hypersurfaces becomes
\be
{\v(\sigma+\delta\sigma,\Phi+\delta\Phi)\over\v(\sigma,\Phi)}
\approx \biggl[{1+(\sigma+\delta\sigma)^2/4n^2
\over 1+\sigma^2/4n^2}\biggr]^{1/2}\,\biggl(1+{\delta\Phi\over\Phi}
\biggr)^{\gamma_1/2},
\ee
and therefore 
\be \label{typicalI}
\langle\delta\sigma\rangle\approx {\sigma\,H^2\over 16\pi^2 n^2},
\ee
\be  \label{typicalII}
\langle\delta\Phi\rangle\approx \Phi.
\ee
In order to compute the spectrum of fluctuations 
we use the standard result for the adiabatic energy perturbations 
in a CDM-dominated universe \cite{lyth,BD2} 
\be
\biggl\langle{\delta\rho\over\rho}\biggr\rangle
= -{6\over 5}H\,\biggl[\dot\sigma\,\delta\sigma+
2\,{(\omega+\Phi\omega')^2\over\omega}\,{\dot\Phi\over\Phi}
\,\delta\Phi\biggr]\biggl(\dot\sigma^2+2\,{\omega\over\Sigma^2}
\,{\dot\Phi^2\over\Phi}\biggr)^{-1}, 
\ee
where we have transformed 
the standard notation to the hyperextended inflation formalism. 
Taking into account the approximation $\Sigma\approx\omega$ in 
(\ref{toyI}) and the results (\ref{typicalI})(\ref{typicalII}) we 
get 
\be
\biggl\langle{\delta\rho\over\rho}\biggr\rangle\approx
{\sigma^2\over 20n^2}\,{H^2\over (2\pi)^2\,\Phi},
\ee
which is to be evaluated at $N\approx 65$ $e$-foldings after 
inflation. The dependence on $\sigma$ can be eliminated via 
(\ref{conservation2}) and is dominated by a constant term dependent 
on the initial fields, i.e.
\be 
\biggl\langle{\delta\rho\over\rho}\biggr\rangle\approx
{1\over 20\pi^2}\,\biggl({\Phi_0\over\Phi}\biggr)\,H^2,
\ee
so essentially if we consider (as we have in the previous sections), 
that the predominant value of $\Phi$ in a physical volume of the 
universe of the horizon scale is $\Phi_{\rm max}$ (so that 
$M_P(\Phi_{\rm max})\sim 10^{19}$ GeV)  
then the typical density contrast becomes 
$\langle\delta\rho/\rho\rangle \sim \Phi_0/\Phi_{\rm max}$. Therefore 
if one is to adjust this to the astrophysical constraint 
$\langle\delta\rho/\rho\rangle\lsim 10^{-4}$, then 
$\Phi_{\rm max}\sim 10^{18}$ GeV is only compatible with a sufficient 
period of evolution, so that a much smaller initial value 
$\Phi_0\sim 10^{13}-10^{14}$ GeV can grow in several orders of 
magnitude to reach $\Phi_{\rm max}$ at the end of inflation. This entails 
at the same time, within the model $\omega\approx \eta/\Phi$ a 
decrease of the same order of magnitude in the value of $\omega$ 
with respect to its initial value.

\section{How typical is GR?}

The notion of something being {\it typical}, in cosmology as 
in any other field, is defined within the  
context of an ensemble. In order to address the question 
of whether an object, quantity or phenomenon is typical or 
not we need to have a knowledge of the entire set or phase 
space that is accessible to us with the physics and initial 
conditions we set out with. In theories of hyperextended 
inflation, the larger ensemble 
that sets our standard of reference to define what is 
typical is the universe as a whole. \\
\\
In a quantum cosmological model that is governed by the 
hyperextended inflation action (\ref{action}) the ensemble 
we need to consider is the physical space that results from the 
evolution after an arbitrary lapse of time. We know  
from the principle of stationarity that the properties 
of the inflating and non-inflating regions, volume ratios, etc, 
swiftly reach asymptotic values, and therefore we can 
safely investigate the likelihood of physical quantities 
regardless of the age of the universe. In fact, the 
principle of stationarity enables us to extend the evolution 
to the infinite past and to think of the notion of what 
is typical in an universe of arbitrary size. Everything 
is this model is ultimately determined by the choice of 
a potential; a choice that is made ad hoc or at best motivated 
by particle physics and indirect constraints from structure 
formation models. The potential determines the global 
structure of the universe, the ratio of inflating regions 
and non-inflating ones, it influences the resulting 
spectrum of fluctuations at the end of inflation, and it 
determines the distribution of values of the Planck mass 
as well as the coupling $\omega$ via the equations 
(\ref{SRR1})-(\ref{SRR3}) that we have investigated in \S 3.  
In general, the choice of potential determines whether 
a given physical quantity is typical or not. 

\subsection{Likelihood, naturalness and typical quantities} 

Let us agree then on the convention that a physical quantity 
is {\it typical} when it is most probable within a universe that is  
governed by hyperextended inflation. In this Section we 
would like to address the question of whether GR is typical 
in this framework. GR is in essence reproduced by 
BD gravity in the limit of large $\omega$, and therefore 
we ought to look at how typical this situation is. The regions 
that are still undergoing inflation are of no direct relevance 
to this discussion and we are interested in the values of 
the fields at the EoI boundary, after which $M_P$ and $\omega$ 
remain essentially constant. Therefore we examine  
regions that have thermalized or are located at the neighbourhood 
of the EoI, to see if a typical region of this kind 
is compatible with GR.\\
\\
In \S 3 we have seen that a potential with 
a finite value of $\Phi_{\rm max}$ is the only likely framework 
to reproduce GR, and $\Phi_{\rm max}\to\infty$ entails that 
the likeliest value of $\omega$ is $-1$ as can be seen from 
(\ref{phi-astro}). Theories that yield a finite $\Phi_{\rm max}$ 
given by (\ref{hyp-phimax}) are to a good approximation well 
described by the Ansatz 
(\ref{toyI}) near the EoI boundary and lead to the likeliest 
value $\omega_*\approx \eta/\Phi_{\rm max}$, and $\eta$ is 
determined by the initial conditions. In fact, as one can 
see from (\ref{toy-ratioI})(\ref{toy-ratioII}), the probability 
distribution for $\omega$ is not extremely steep, it is typically 
a powerlaw and therefore, strictly speaking, a typical region of 
the universe has a value 
of $\Phi$ that is spread over a small range within the sharp wedge 
between the curves EoI and BoI (thick solid and dotted curves 
respectively) in 
Fig.~1. The values of $\Phi$ that fall within this neighbourhood 
$\lsim\Phi_{\rm max}\sim 10^{18}$ GeV 
lead to values of $\omega$ that are, to a certain 
extent, also typical. An appropriate choice of these parameters 
such that $\omega_*$ satisfies the rather conservative constraint 
$\omega\gsim 500$ would therefore make GR a typical theory. \\
\\
The next question arises as to whether one should allow a greater 
freedom for a choice of potential, or whether it is physically 
{\it natural} to choose one, 
and to constrain the parameters conveniently 
so that one derives a result that tells us that GR is a typical 
theory. It could be argued that a potential ought to be picked 
out for its likelihood by the dynamics, because it {\it prevails} 
as a typical theory in a framework that allows all possible 
potentials, rather than by starting out with an ad hoc choice. 
In principle, one can undertake this step further and envisage 
a quantum cosmological model where all possible inflation potentials 
find a realization, such as in Fig.~3. The universe is then 
subdivided in an infinite number of regions $v_i$ or 
{\it subuniverses} that are characterized by a given potential 
$V_i(\sigma)$. An arbitrary region $\v_i$ of the universe is 
therefore totally equivalent to the hyperextended theory that we have 
investigated. The universe as a whole however is not and it is 
not correctly described by the likelihood ratios 
(\ref{hyp-ratio}) that we have computed 
for a powerlaw potential. To find the correct likelihood ratios 
we need to integrate the volumes of 
the homogeneous hypersurfaces $\v(\sigma,\Phi)$ over the entire 
space $\cal M$ of potentials, i.e. $\tilde\v(\sigma,\Phi)\equiv 
\int_{\cal M}\d V\;\v(\sigma,\Phi)$. It is easy to 
see that quantities that may be typical within a subuniverse 
$\v_i$ may be not only not typical within the larger ensemble 
$\{\v_j\}$, but also probably highly unlikely.\\
\\
It is quite apparent that investigating whether GR is typical 
is not without assumptions, some of which can be conflicting. 
We can classify the following two sets of inequivalent assumptions:

\newcounter{bean}
\begin{list}{\Roman{bean}}{\usecounter{bean}}
\item Assume GR is typical, and therefore the possible 
inflaton potentials are restricted to the class $\Phi_{\rm max}<\infty$ 
and the free parameters are of the adequate order of magnitude 
so that $\omega_*\approx \eta/\Phi_{\rm max}\gsim 500$. 
\item Assume a certain inflaton potential(s), either from 
particle physics or reconstruction of the potential, and therefore we 
can conclude whether GR is typical or not depending on whether 
$\Phi_{\rm max}<\infty$ or $\Phi_{\rm max}\to \infty$ and the 
value of $\omega(\Phi_{\rm max})$. 
\end{list}

\noindent Both approaches are clearly inequivalent and it can 
be argued that they can be motivated for different reasons. 
Assumption (I) is based on the so-called {\it principle 
of mediocrity} \cite{mediocrity}, which states briefly 
that the physical quantities 
observed in our neighbourhood of the universe are typical quantities, 
given that there is nothing especial about the region of the universe 
we inhabit with respect to other regions. Therefore our immediate 
neighbourhood cannot be singled out as a region that is untypical 
in any way, and GR has to be a perfectly typical theory.
\footnote{It must be noted that, strictly speaking, the principle of 
mediocrity is not a good name for this notion, since {\it mediocrity} 
not only entails being average and unremarkable but, rather, below 
the average and under-achieving. There is no reason to choose such 
negative connotations for a principle that is meant to denote what 
is likely, average and typical.} 
In this context, the immediate implication is that the 
class of potentials that would make this possible is restricted 
by the conditions $\Phi_{\rm max}<\infty$ and $\omega_{\rm typical}
\gsim 500$. Therefore, the quantum cosmological scenario that we 
briefly summarized above and depicted in Fig.~3 would be somewhat 
difficult to sustain under the assumption (I). Such a model would 
require that not all potentials that are possible are realized with 
equal probability but that those that satisfy the restrictions imposed 
by (I) are far more probable and prevail. \\
\\
Assumption (II) involves coming up with a potential, that is 
motivated by observations or particle physics. For example, a 
potential that is a quasi-powerlaw with a kink on its slope that 
results in an open inflation theory is, it can be argued, well 
motivated by observations. The question of whether the potential 
chosen yields the appropriate volume ratios like those that we have 
computed in \S 3, in order to conclude that GR is a typical theory 
is a different matter. One needs to check this by computing 
the distribution of $\omega$ for each particular potential. 
In some cases GR may be a typical theory, in others it may not.

\subsection{Is there a typical inflaton potential?}

Following the analysis in \S 3, where we have based the notion 
of likelihood on the physical volume occupied by a given hypersurface, 
we can extend this notion to investigate how typical some potentials 
are with respect to others. Certainly, the potentials that yield 
$\Phi_{\rm max}\to\infty$ will be the likeliest ones, as they do by 
far occupy the largest physical volume (although the number of 
potentials that yield $\Phi_{\rm max}<\infty$ is far greater). 
In these theories, the fields can grow without limit during 
the course of inflation and occupy an arbitrarily large volume. 
Amongst this class of potentials, the 
relative likelihood of two potentials is difficult to quantify. 
Within the class of potentials that yields a finite $\Phi_{\rm max}$, 
there are powerlaw potentials ($n>2$), exponential potentials 
or double-well potentials. For powerlaw potentials, 
we have seen from Fig.~1 that there are a fair amount of solutions, 
as an arbitrary $n>2$ gives a finite $\Phi_{\rm max}$, whereas 
only $n=1,2$ yield $\Phi_{\rm max}\to\infty$. \\
\\
If we adopt assumption (I) of \S 4.1, i.e. that GR is typical, 
then we conclude that the powerlaw potentials that have the 
smallest values of $\lambda$ are favoured and are the likeliest, 
as we have from (\ref{hyp-phimax}) that typically 
$\Phi_{\rm max}\sim\lambda^{-1/(n-2)}$. Therefore, a small value 
of $\lambda$ such as (\ref{lambda}) would be perfectly consistent 
with GR being a typical theory. According to this argument, hence 
even though particle physics does not provide a mechanism to discriminate 
values of $\lambda$, and all possible values of $\lambda$ are 
feasible with equal probability, only the smallest ones turn out to 
be naturally the likeliest ones because they yield the largest 
$\Phi_{\rm max}$, and thus this creates an ideal scenario to 
accept that GR is a typical theory. Even though GR is indeed a 
typical theory within the ensemble of regions governed by powerlaw 
potentials, or at any rate not an unlikely one (using a more 
conservative viewpoint), GR is not a typical theory within the larger 
ensemble of all possible potentials, as these are donimated by 
$\Phi_{\rm max}\to\infty$ regions. These include a number of exotic 
potentials, such as logarithmic, etc. On the other hand, if we were 
to exclude from all possible realizations those potentials that 
are remotely exotic, and one just allowed a physical realization to 
a more conventional class of potentials, say e.g. powerlaw, exponential 
and double-well potentials, then within this restricted subclass 
$\Phi_{\rm max}<\infty$ is a predominant solution and we are back to 
the assertion that GR is a typical theory. \\
\\
It could be said that this is a rather contrived way of looking for 
a typical potential or for discarding potentials that modify the 
notion of what is typical in a way that we do not want. 
On the other hand, if we abandon the question of 
looking for a typical potential that leads to GR as the likeliest 
theory, and we allowed all possible potentials to find a realization, 
as we described in the previous section, we find that GR is an extremely 
unlikely and untypical theory. However one could argue that this might 
not be all that discouraging. From the anthropic point of view, this 
could be perfectly acceptable as an untypical configuration on which 
the very existence of human life and its relationship to the physical 
world relies. This viewpoint belongs to the class of assumptions (II).
From what we have seen, it is apparent that adopting either (I) or 
(II) can lead to completely different notion of how typical GR is. 

\section{Conclusions}

We have investigated some aspects of extended and hyperextended 
inflation that are related to the question of how typical GR is. 
Two important results are the derivation of the conservation laws 
(\ref{conservation})(\ref{conservation2}), which give us the 
classical trajectories of the fields between the boundaries BoI and 
EoI. The quantum fluctuations of the fields, due to the stochastic 
nature of inflation, are superimposed on the classical trajectories, 
and as we have discussed in \S 3, these contribute to a significant 
extent to drive the fields to larger values. It is fundamental to 
take this effect into account in computing the likelihood of a 
observer being in a region ($\sigma$,$\Phi$)=const, that is proportional 
to the physical volume $\v(\sigma,\Phi)$ of the corresponding homogeneous 
hypersurface. \\
\\
We have also divided inflaton potentials in two classes: 
\begin{itemize}
 \item [A] The BoI and EoI curves intersect, therefore 
there is a maximum value of the BD field $\Phi_{\rm max}<\infty$ 
for which inflation can still take place,
\item [B] BoI and EoI do not intersect, and therefore 
$\Phi_{\rm max}\to\infty$. 
\end{itemize}
\noindent The first class of potentials is of 
interest and has been employed in most calculations in \S 3, 
whereas the second class is problematic in that the fields 
grow to arbitrarily large values and the most typical scenario is 
one of $\omega= -1$ and a departure from this value takes place with 
negligible probability. These so-called "run-away" solutions 
are discussed in \S 2.1. Fortunately, most powerlaw potentials 
($n>2$) belong to class A, and our analysis of hyperextended 
inflation in \S 3 has been applied to these. It is straightforward 
to derive the same results for arbitrary potentials, of either 
class A or B, as the main equations presented are valid for generic 
potentials, provided the slow-roll approximation applies. \\
\\
In \S 3 we have investigated hyperextended inflation theories making 
an Ansatz for $\omega(\Phi)$ based on the 
principle of least coupling. The coefficients $\eta_i$ in (\ref{w-taylor}) 
that determine $\omega$ are given by loop corrections of the lowest-order 
effective action in string theory, and these are not easy to 
calculate. However, for the toy models that we have considered in 
\S 3.3--4, one can constrain the coefficients in terms of the 
initial conditions, and in general, in terms of the approximate 
behaviour in the vicinity of EoI as described in \S 3.4. Other constraints 
on $\eta_i$ can be derived for example from the bubble size spectrum. 
As is well known, extended inflation suffered from the so-called 
{\it big-bubble problem} (see e.g. last reference in \cite{constraints}). 
Hyperextended inflation enables us 
to adjust the parametrization of $\omega$ and fit $\eta_i$ 
as best as possible to avoid this problem in particular. \\
\\
In \S 3 we have seen that powerlaw potentials of class A are 
to a large extent consistent with GR being a typical theory 
whereas for those in class B, GR is very untypical.  
In \S 4 we have discussed the conceptual framework of a quantum 
cosmological framework that is governed by hyperextended inflation. 
A scenario that allows a realization of all possible scalar potentials 
results in an ensemble of subuniverses $\v_i$ each one of which is 
described by the distributions computed in \S 3. In this senario 
GR is not a typical theory as most of the volume is dominated by 
regions governed by potentials of class B. In a more restricted 
scenario that only allows a realization for potentials of class A 
(that also happen to be the less exotic potentials and physically 
more realistic), it is consistent that GR is a typical theory. \\
\\
In conclusion, it must be said that the notion of how typical 
GR is, as presented in the title of this paper, is a question that 
escapes a straightforward answer. As we have discussed in \S 4, the 
two different assumptions (I) and (II) can lead us to reach a 
different conclusions, and the uncertainty derives from our lack 
of knowledge of the inflaton potential. However many things can 
be said about whether GR is typical or untypical depending on the 
potential chosen. The scenario that allows a realization of 
all possible potentials $\Phi_{\rm max}<\infty$ is a particularly 
promising framework, that results in GR being a typical theory.

\section*{Appendix. Volume ratios in extended inflation} 

In this appendix we work out the details of the derivation of 
(\ref{ratio}). From (\ref{jsigma})(\ref{jphi}) we have 
\be \label{jsigma2}
J_{\sigma}\approx -2\Phi\,\biggl({V'\over V}\biggr)\,\calP_c,
\ee
\be \label{jphi2} 
J_{\Phi}\approx 2\,\biggl({\Phi\over\omega}\biggr)\,\calP_c,
\ee
where $\calP_c$ is given by (\ref{pc}). As was shown in 
\cite{BD2}, the comoving probability $\calP_c$ is 
strongly dependent of the choice of time variable. It is 
desirable thus to use a measure of the likelihood of 
the values of the fields that is independent of the choice 
of time parameter. The principle of stationarity tells 
us that, given that a sufficiently long period of inflation has 
elapsed, the volume of a hypersurface $\v(\sigma,\Phi)$  
becomes a constant fraction of the 
total volume of the universe. Of course 
the total volume of the universe $\v_T$ grows indefinitely and the 
volumes of homogeneous hypersurfaces are also unbounded, but 
the volume ratio of two homogeneous hypersurfaces or that of 
a homogeneous hypersurface with respect to the total volume 
of the universe remains finite. Throughout this article 
we have arbitrarily chosen that the measure of likelihood 
of the fields ($\sigma$,$\Phi$) is given by the ratio of 
the volume of the hypersurface ($\sigma$,$\Phi$)=const 
to the volume occupied by thermalized regions:
\be \label{r}
r={\v(\sigma,\Phi)\over\v_*},
\ee
where 
\be \label{volume1}
\v(\sigma,\Phi)= \v_0\;\biggl|\int_0^{t_c}\!\d t\,\e^{3t}
\,(\b J\cdot\hat{l})\biggr|,
\ee
\be \label{volume2}
\v_*=\v_0\;\biggl|\int_0^{t_c}\!\d t\,\e^{3t}
\int_{\rm EoI} \!\d l\,(\b J\cdot\hat{n})\biggr|,
\ee
where $\v_0$ is an arbitrary volume of an initial homogeneous 
hypersurface, $\hat l$ is a tangent vector to the curve 
(\ref{conservation}) at ($\sigma$,$\Phi$), $\hat n$ is a normal 
vector to the EoI boundary (\ref{EoI}). The parameter 
$t_c$ is an arbitrarily large {\it cut-off time} that regularizes 
the volumes (\ref{volume1})(\ref{volume2}). The volume ratio 
(\ref{r}) is independent of $t_c$ as shown in 
\cite{method,BD-inv-perturbations} and therefore we take the 
limit $t_c\to\infty$. \\
\\
After some algebra it is easy to show that in the case of a 
powelaw potential 
\be \label{volume12}
\v(\sigma,\Phi)_{\rm regularized}= 2\v_0\,\biggl\{{\exp\Big[(3-
\gamma_1)\,t_c\Big]-1\over 3-\gamma_1}\biggr\}\,\Phi^{{\omega\over 2}
\gamma_1}\,\biggl(1+{\sigma^2\over 4n^2\,\omega^2}\biggr)^{1/2},
\ee 
\be  \label{volume22} 
\v^*_{\rm regularized}=\v_0\, \biggl\{{\exp\Big[(3-
\gamma_1)\,t_c\Big]-1\over 3-\gamma_1}\biggr\}\, 
{4\over\omega\gamma_1}
\,\Phi_{\rm max}^{{\omega\over 2}\gamma_1+1}.
\ee
Therefore the volume ratio is 
\be \label{fullratio}
r(\sigma,\Phi)= {\omega\over2}\,\gamma_1\,
\Phi_{\rm max}^{-{\omega\over 2}\,\gamma_1-1}\,
\biggl(1+{\sigma^2\over 4n^2\,\omega^2}\biggr)^{1/2}\,
\Phi^{{\omega\over 2}\,\gamma_1}
\ee
which is the result that has been used in (\ref{ratio}) as 
a measure of the likelihood of a given configuration ($\sigma$,$\Phi$). 
The presence of $\Phi_{\rm max}$ arises from the integral over the 
EoI boundary in (\ref{volume2}). This quantity, like $\gamma_1$ is 
model-dependent. For powerlaw potentials $n\leq 2$ we 
have seen however in \S 2 that $\Phi_{\rm max}\to\infty$. According to 
(\ref{fullratio}) in that limit only the fields at infinity would yield 
a non-vanishing value of $r$, but we rule out that possibility as 
it would make our region of the universe an extraordinarily unlikely one.

\newpage

\begin{figure}
\vspace*{14cm}
\includegraphics{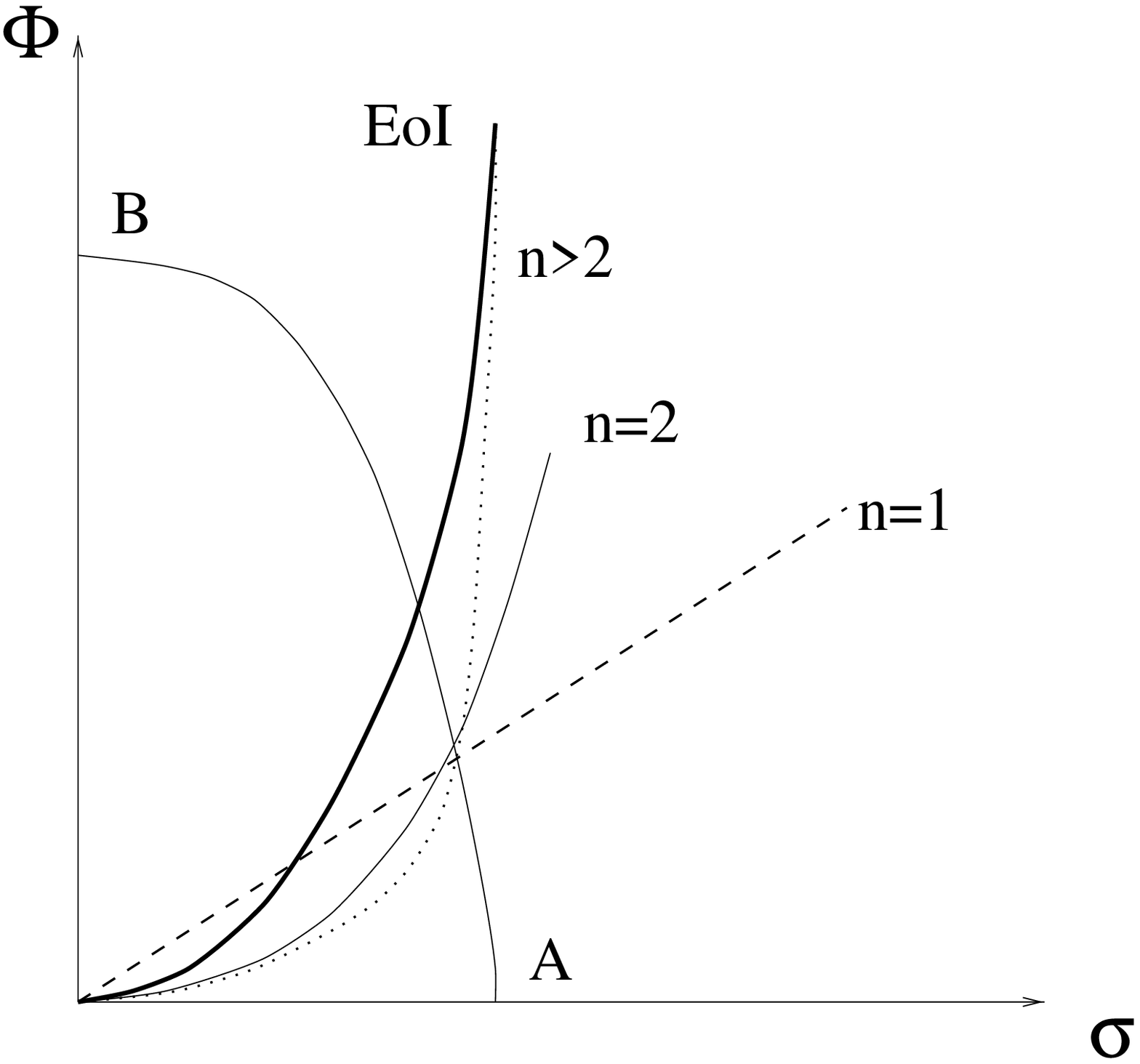}
\caption{Classical evolution of the fields for a powerlaw 
potential. The solid curve $AB$ represents the parabola 
(4). Given some initial conditions ($\sigma_0$,$\Phi_0$) 
on $AB$, the fields roll along this curve in the direction $A\to B$, 
and quantum jumps take the fields from one classical trajectory 
to another. The thick solid line represents the EoI boundary. 
The BoI boundary is represented by: the dashed line for 
$n=1$, solid line for $n=2$ and dotted line for $n>2$. EoI and 
BoI intersect at $\Phi_{\rm max}$ for $n>2$.}
\label{figroll}
\end{figure}

\newpage

\begin{figure}
\vspace*{14cm}
\includegraphics{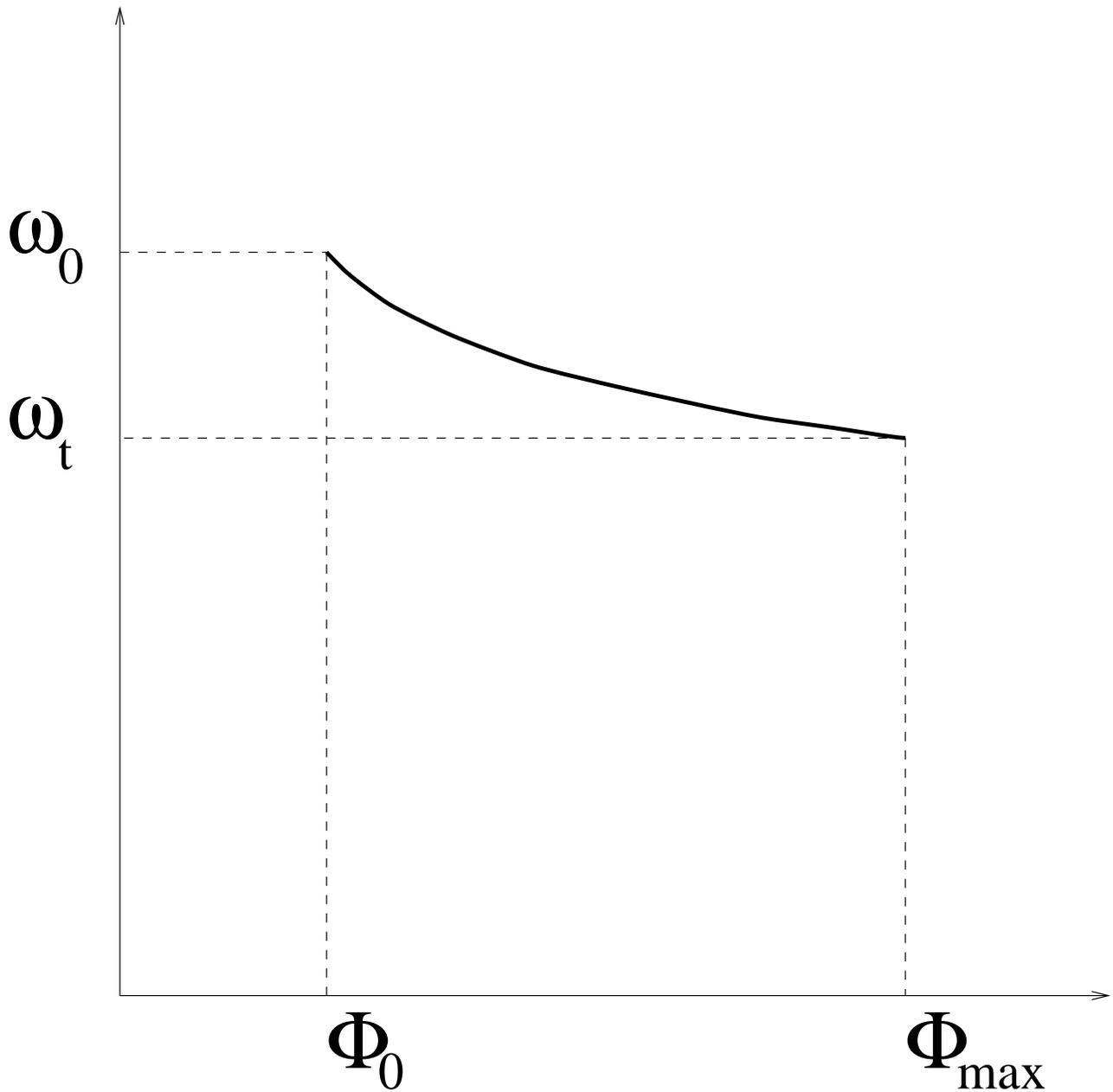}
\caption{Evolution of $\omega$ for the toy model 
$\omega\sim\eta /\Phi$. The value of $\omega$ decreases 
during the course of inflation and becomes constant after 
crossing the EoI boundary. The lowest value of $\omega$ 
possible corresponds to $\Phi_{\rm max}$, the highest value 
of $\Phi$ along the EoI for which inflation still occurs. This 
is at the same time the likeliest or most {\it typical} value 
of $\omega$.} 
\label{toyfig}
\end{figure}

\newpage

\begin{figure}
\vspace*{14cm}
\includegraphics{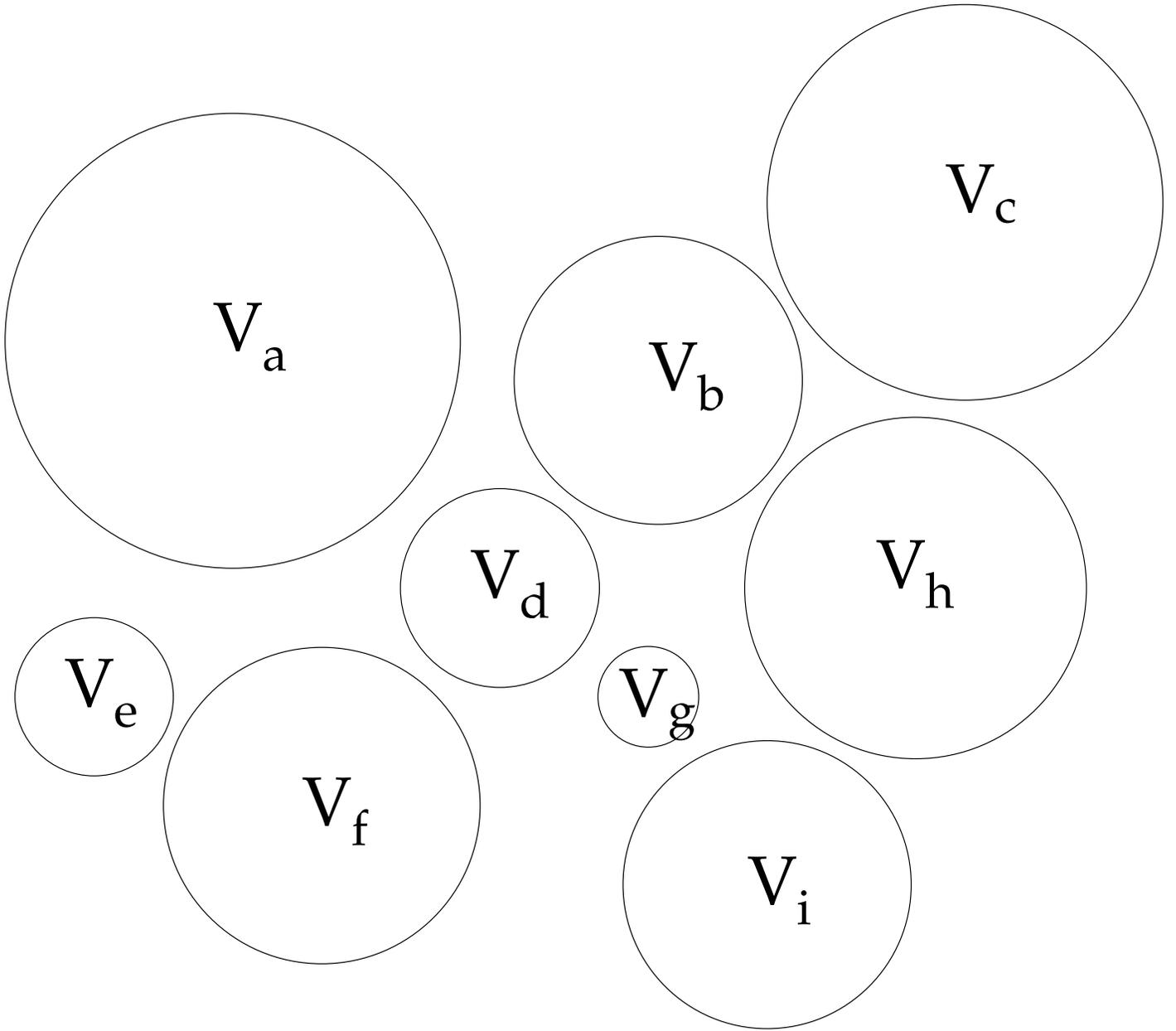}
\caption{The hyperextended inflationary universe. An arbitrary 
region $\v_i$ is governed by a potential $V_i(\sigma)$ via 
the hyperextended inflation action (1). All possible potentials 
find a realization and are equally probable. The probability 
distributions of the fields and the volume ratios of homogeneous 
hypersurfaces within $\v_i$ are given by the equations of \S 3. 
Some of the $\v_i$ occupy a larger fraction of the entire volume 
of the universe than others, depending on the relative magnitude 
of $\Phi_{\rm max}^{(i)}$. An observer is naturally typically 
located in a region $\Phi_{\rm max}\to\infty$ as these regions 
span by far the largest fraction of the volume.} 
\label{potfig}
\end{figure}

\end{document}